# Large and negative Kerr nonlinearity in PdSe$_2$ dichalcogenide 2D films

Linnan Jia,[1] Jiayang Wu,[1] Tieshan Yang,[2] Baohua Jia,[1,2] David J. Moss[1]*

[1]*Optical Science Centre, Swinburne University of Technology, Hawthorn, VIC 3122, Australia*

[2]*Centre for Translational Atomaterials, Swinburne University of Technology, Hawthorn, VIC 3122, Australia*

*\* E-mail: dmoss@swin.edu.au*

## Abstract

We report a large third-order nonlinear optical response of palladium diselenide (PdSe$_2$) films – a two-dimensional (2D) noble metal dichalcogenide material. Both open-aperture (OA) and closed-aperture (CA) Z-scan measurements are performed with a femtosecond pulsed laser at 800 nm to investigate the nonlinear absorption and nonlinear refraction, respectively. In the OA experiment, we observe optical limiting behaviour originating from large two photo absorption (TPA) in the PdSe$_2$ film of $\beta = 3.26 \times 10^{-8}$ m/W. In the CA experiment, we measure a peak-valley response corresponding to a large and negative Kerr nonlinearity of $n_2 = -1.33 \times 10^{-15}$ m$^2$/W – two orders of magnitude larger than bulk silicon. We also characterize the variation of $n_2$ as a function of laser intensity, observing that $n_2$ decreases in magnitude with incident laser intensity, becoming saturated at $n_2 = -9.96 \times 10^{-16}$ m$^2$/W at high intensities. These results verify the large third-order nonlinear optical response of 2D PdSe$_2$ as well as its strong potential for high performance nonlinear photonic devices.

## 1. Introduction

With their extraordinary nonlinear optical properties, two-dimensional (2D) layered materials such as graphene [1-4], graphene oxide (GO) [5-9], transition metal dichalcogenides (TMDCs) [10-12], and black phosphorus (BP) [13-15] have attracted a great deal of interest, enabling diverse nonlinear photonic devices with vastly superior performance compared to bulk materials. Amongst them, TMDCs (MX$_2$, M = transition metal and X = chalcogen), with bandgaps in the near infrared to the visible region, have opened up promising new avenues for photonic and optoelectronic devices. [2, 16]. Previously, the prominent nonlinear optical properties of TMDCs such as MoS$_2$ [10, 17, 18], WS$_2$ [18-20] and MoSe$_2$ [21, 22], including a high Kerr nonlinearity and strong saturable absorption (SA), have been reported. These have provided the basis of many nonlinear photonic devices for applications where, for example, a few mono-layers of MoS$_2$ and WS$_2$ have been used as broadband, fast-recovery saturable absorbers for mode locking in pulsed fiber lasers [10, 23, 24]. In addition, nonlinear optical modulators and polarization dependent all-optical switching devices have been realized based on ReSe$_2$ [25] and SnSe [26].

Palladium diselenide (PdSe$_2$), a new 2D noble metal dichalcogenide in the TMDC family, has recently attracted significant interest [27-31]. Similar to the puckered structure of BP, it has a puckered pentagonal atomic structure − with one Pd atom bonding to four Se atoms and two adjacent Se atoms covalently bonding with each other. This low-symmetry structure makes PdSe$_2$ possess unique in-plane anisotropic optical and electronic properties [27, 29], featuring an in-plane non-centrosymmetric structure, in contrast to its cousin PtSe$_2$. Further, the bandgap of PdSe$_2$ is layer-dependent, varying from 0 eV (bulk) to 1.3 eV (monolayer) - a property well suited for photonic and optoelectronic applications – in particular, for wavelength tuneable devices [27, 29, 31]. Moreover, PdSe$_2$ is highly air-stable, indicating its robustness and potential for practical applications as compared with BP. The high carrier mobility and anisotropic Raman spectroscopy of 2D PdSe$_2$ layers have been investigated [27, 30] as well as highly-sensitive photodetectors from the visible to mid-infrared wavelengths [31, 32]. Recently, the optical nonlinear absorption of PdSe$_2$ nanosheets has also been reported in the context of mode-locked laser applications [33-35]. To date, however, its optical Kerr nonlinearity has not been investigated.

Here, we report measurements of the third-order nonlinear optical response of 2D PdSe$_2$ films. Experimental results using the Z-scan technique with femtosecond optical pulses at 800 nm show that PdSe$_2$ films exhibit a large and negative (self-defocusing) Kerr nonlinearity ($n_2$) of ~ -1.33×10$^{-15}$ m$^2$/W, which is two orders of magnitude larger than bulk silicon. Further, we measure a large nonlinear absorption $\beta$ of ~ 3.26 ×10$^{-8}$ m/W, which originates from TPA in the PdSe$_2$ films. In addition, we investigate the intensity dependence of the nonlinear response of PdSe$_2$, finding that the absolute magnitude of the Kerr nonlinearity $n_2$ initially decreases slightly with incident laser intensity, becoming saturated at higher intensities. Our results show that the extraordinary third-order nonlinear optical properties of PdSe$_2$ have strong potential for high-performance nonlinear photonic devices.

## 2. Material preparation and characterization

Figure 1(a) illustrates the atomic structure of PdSe$_2$ crystals. The unique puckered pentagonal structure is different to other TMDCs like MoS$_2$ and WS$_2$. The Se-Pd-Se layers stack with weak van der Waals interactions to form a layered structure [27, 36, 37]. In each monolayer, the pentagonal rings are formed with one Pd atom bonding to four Se atoms and two adjacent Se atoms covalently bonding with each other, which is similar to the puckered structure of BP, and yields both anisotropic and non-centrosymmetric properties of PdSe$_2$ [27, 37]. More importantly, unlike the rapid degradation of BP under ambient conditions, PdSe$_2$ has significantly better air-stability [29-31]. Together, these properties of PdSe$_2$ make it promising for high performance photonic and optoelectronic applications.

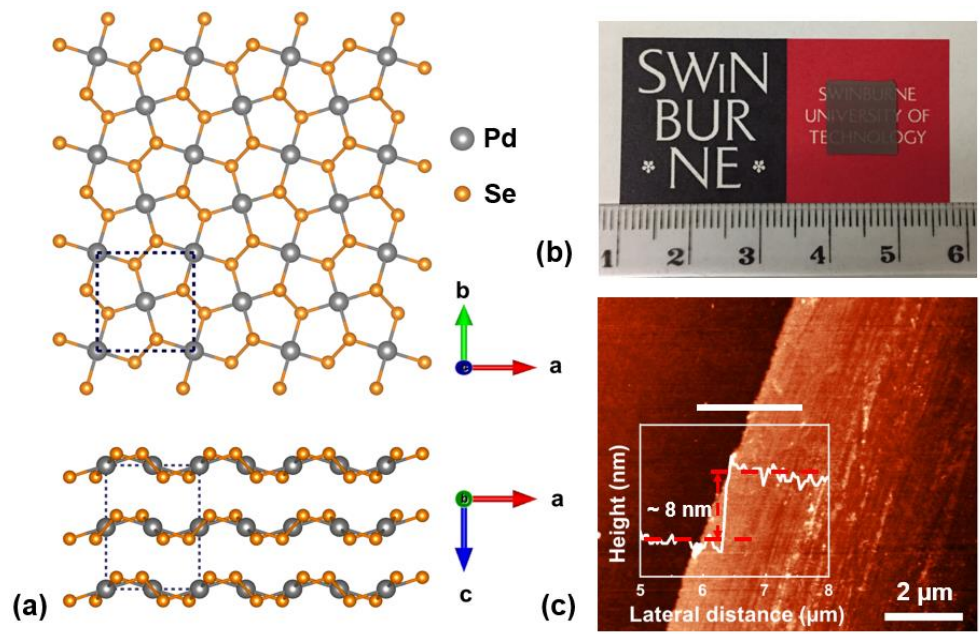

**FIG. 1.** (a) Crystal structure of PdSe$_2$. (b) Photograph of prepared multilayer PdSe$_2$ film. The unit for the numbers on the ruler is centimeter. (c) AFM height profile of the multilayer PdSe$_2$ film.

Here, we investigated large-area multilayer PdSe$_2$ films deposited on transparent sapphire substrates. The PdSe$_2$ films were synthesized via Chemical vapor deposition (CVD) [30, 38]. A three-zone tube furnace was used to grow the PdSe$_2$ films with palladium chloride (PdCl$_2$) and selenium (Se) as precursors. Se and PdCl$_2$ powders were placed at Zone 1 with a heating temperature of 250 °C and Zone 2 heated up to 500 °C, respectively. The evaporated Se and Pd precursors were then transported by the carrier gas of Ar/H$_2$ to Zone 3 at a high temperature of 600 °C, in which the continuous PdSe$_2$ films were synthesized on an atomically flat sapphire substrate. The films were polycrystalline, as is typical for CVD synthesized films, with crystal sizes varying from 10's of nanometres up to 100 nm. Because of the polycrystalline nature of the films, the inversion symmetry breaking properties (i.e., non-centrosymmetric) of the single PdSe$_2$ crystals could not be observed on optical wavelength scales in the macroscopic PdSe$_2$ continuous films studied in this work. Figure 1(b) shows a photograph of the prepared sample, indicating a high uniformity over the whole substrate. The morphology image and height profile of the prepared PdSe$_2$ films were characterized by atomic force microscopy (AFM), as illustrated in Fig. 1(c). The film thickness was measured to be ~ 8 nm, which corresponds to ~20 layers of PdSe$_2$ [27, 30, 36].

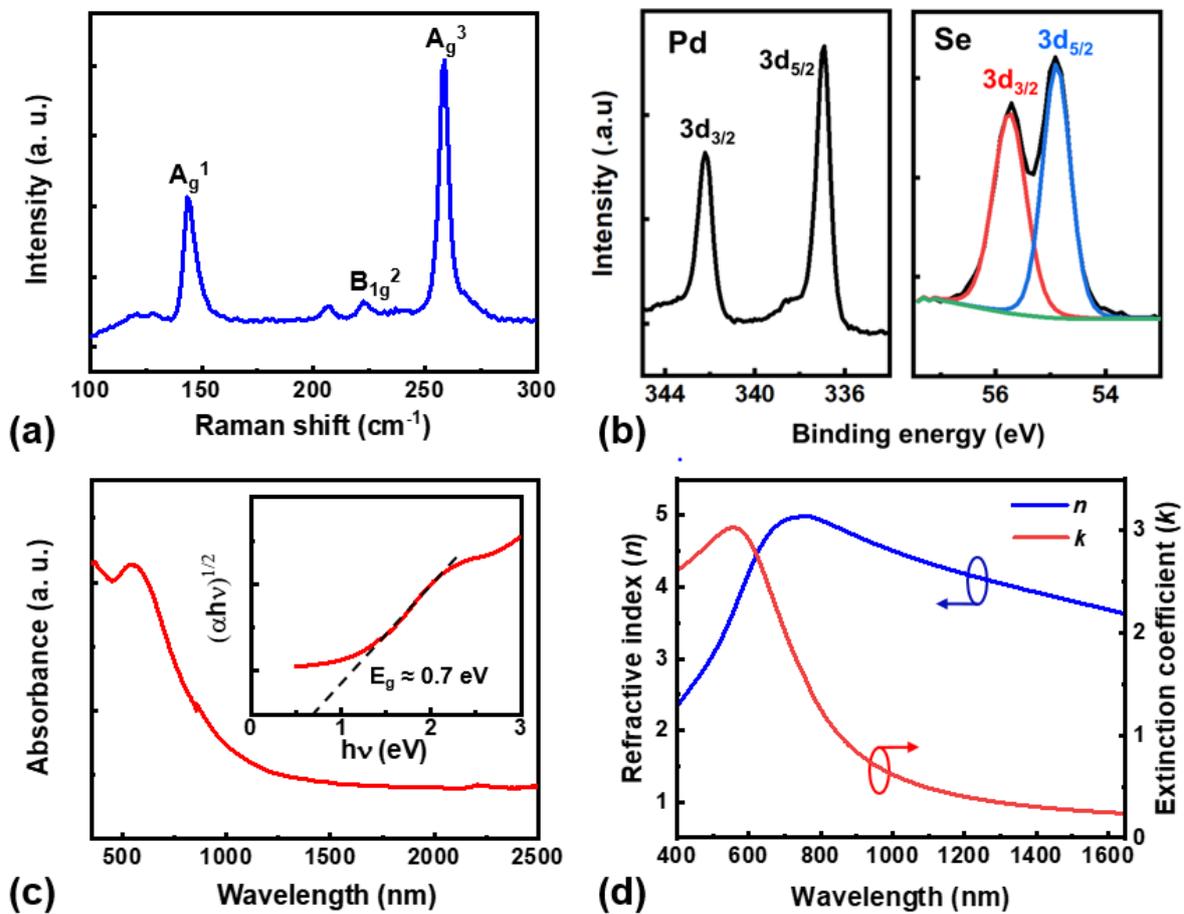

**FIG. 2.** Characterization of the prepared PdSe$_2$ film. (a) Raman spectrum excited via a 514-nm laser. (b) X-ray photoelectron spectroscopy (XPS) spectra. (c) UV-vis absorption spectrum. Inset shows the extracted Tauc plot. (d) Measured refractive index ($n$) and extinction coefficient ($k$).

Figure 2(a) shows the Raman spectrum of the prepared PdSe$_2$ film excited with a laser at 514 nm. Three representative phonon modes are observed, including the A$_g^1$ (~145.5 cm$^{-1}$) and B$_{1g}^2$ (~222.5 cm$^{-1}$) vibrational modes that correspond to the movement of Se atoms and the A$_g^3$ (~ 258.8 cm$^{-1}$) mode that relates to the relative movements between Pd and Se atoms [27, 30, 32]. To further characterize the film quality, X-ray photoelectron spectroscopy (XPS) was employed to measure the binding energy of PdSe$_2$. Figure 2 (b) shows the XPS results of Pd 3d and Se 3d core levels for the PdSe$_2$. The peaks of the fit at ~ 342.2 eV and ~ 336.9 eV are attributed to the Pd $3d_{3/2}$ and Pd $3d_{5/2}$, respectively, whereas the peaks at ~ 55.7 eV and ~ 54.9 eV correspond to Se $3d_{3/2}$ and $3d_{5/2}$, respectively. These results are consistent with previous reports [30, 32] and demonstrate the high quality of our prepared $PdSe_2$ films.

To characterize the linear absorption and optical bandgap, the optical absorption spectrum (from 400 nm to 2500 nm) of the $PdSe_2$ film was measured with ultraviolet-visible (UV-vis) spectrometry, as shown in Fig. 2(c). As expected, the linear absorption of the $PdSe_2$ film increased dramatically from the infrared to visible wavelength regime. The optical bandgap of the $PdSe_2$ film can be estimated from a Tauc plot of $(\alpha h\nu)^{1/2}$ versus $h\nu$ based on the Tauc formula, where α and hν represent the optical absorption coefficient and photon energy, respectively.[29, 32, 39] The inset of Fig. 2 (c) shows the Tauc plot extracted from the linear absorption spectrum, where the optical bandgap of the $PdSe_2$ film is estimated to be ~ 0.7 eV and shows good agreement with the reported values for 20 layers of $PdSe_2$ in Refs. [29, 32].

We also characterize the in-plane (TE-polarized) refractive index (*n*) and extinction coefficient (*k*) of the $PdSe_2$ film via spectral ellipsometry, as depicted in Fig. 2(d). For the thin $PdSe_2$ film with a thickness of ~8 nm, the out-of-plane (TM-polarized) response is much weaker. Thus, using spectral ellipsometry we could only measure the in-plane *n* and *k* of the $PdSe_2$ film. It can be seen that the refractive index first increases dramatically with wavelength to reach a peak at ~ 700 nm and then decreases more gradually at longer wavelengths. This trend is similar to that of $PtSe_2$ – another noble metal dichalcogenide of the TMDC family [40]. The extinction coefficient exhibits a significant decrease from 600 nm to 1200 nm, and then the rate of decrease slows down at longer wavelengths. This shows an agreement with the trend of the UV-vis absorption spectrum in Fig. 2 (c) and further confirms the validity of our ellipsometry measurement. These measurements indicate that the films had a bandgap of about 0.7 eV, residing just below the telecommunications wavelength band.

### 3. Z-scan measurement

*3.1 Experimental setup*

To investigate the third-order nonlinear optical properties of $PdSe_2$, we characterized the nonlinear absorption and refraction of the prepared $PdSe_2$ film via the Z-scan technique [14, 41, 42] (Fig. 3), where a femtosecond pulsed laser with a centre wavelength at ~800 nm and pulse duration of ~ 140 fs was used to excite the samples. A half-wave plate combined with a linear polarizer was employed to control the power of the incident light. A beam expansion system consisting of a 25-mm concave lens and 150-mm convex lens was used to expand the light beam, which was then focused by an objective lens (10 ×, 0.25 NA) to achieve a low beam waist with a focal spot size of ~1.6 μm. The prepared samples were oriented perpendicular to the beam axis and translated along the Z-axis with a linear motorized stage. A high-definition charge-coupled-device (CCD) imaging system was used to align the light beam to the target sample. Two photodetectors (PDs) were employed to detect the transmitted light power for the signal and reference arms.

The Z-scan measurements were performed in two stages [14, 41] including an open-aperture (OA) and a closed aperture (CA) measurement. The OA measurement was used to characterize the nonlinear absorption of $PdSe_2$ films where all light transmitted through the sample was collected by a PD. To extract the TPA (*β*) of $PdSe_2$, we fit the measured OA results in Fig. 4(a) with [41, 43]:

$$T_{\text{OA}}(z) \simeq 1 - \frac{1}{2\sqrt{2}} \frac{\beta I_0 L_{\text{eff}}}{\left(1+\left(\frac{z}{z_0}\right)\right)}, \tag{1}$$

where $T_{OA}(z)$ is the normalized optical transmittance; $L_{eff} = (1- e^{-\alpha_0 L})/\alpha_0$ is the effective sample thickness; $\alpha_0$ is the linear absorption coefficient; $L$ is the sample thickness; $I_0$ is the irradiance intensity at the focus; $z$ is the sample position relative to the focus, and $z_0$ is the Rayleigh length of the laser beam.

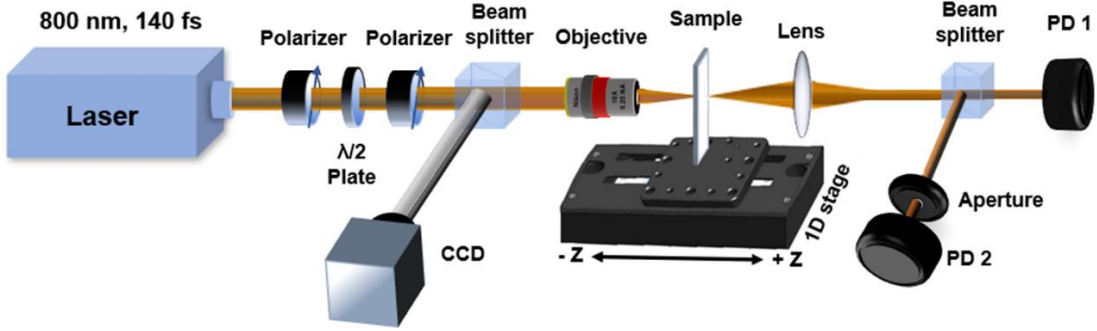

**FIG. 3.** Schematic illustration of Z-scan experimental setup. PD: power detector.

We also characterized the Kerr nonlinearity of the PdSe$_2$ film via the CA Z-scan experiment. In the CA measurement, the change in the optical transmission results from both nonlinear absorption and the nonlinear phase shift induced self-defocusing via the Kerr nonlinearity. The value of $n_2$ for the PdSe$_2$ films at different laser intensities were extracted by fitting the CA measurement with [41, 42]

$$T_{CA}(z, \Delta\Phi_0) = 1 + \frac{4\Delta\Phi_0(\frac{z}{z_0})}{\left[1+\left(\frac{z}{z_0}\right)^2\right]\left[9+\left(\frac{z}{z_0}\right)^2\right]}, \quad (2)$$

where $T_{CA}(z, \Delta\Phi_0)$ is the normalized optical transmittance of the CA measurement and $\Delta\Phi_0 = 2\pi n_2 I_0 L_{eff}/\lambda$ is the nonlinear phase shift, with $\lambda$ denoting the center wavelength of the femtosecond laser. Therefore, the Kerr nonlinearity of the PdSe$_2$ film could be extracted from the ratio of the CA result to the OA result in the usual manner [14, 41, 42].

*3.2. Results and discussion*

We measured the OA curves at different incident laser intensities ranging from 12.08 GW/cm$^2$ to 21.32 GW/cm$^2$. Figure 4(a) shows the OA Z-scan results for the PdSe$_2$ film at three representative intensities. A typical optical limiting behaviour was observed in the OA curves, with the transmission decreasing as the PdSe$_2$ sample was moved through the focal point. We also note that the transmittance dip of the OA curve decreased when the incident laser intensity was increased. Various mechanisms, including nonlinear scattering, reverse saturable absorption (RSA) and two-photon absorption (TPA), may contribute to the observed optical limiting behaviour. The nonlinear scattering mechanism is first excluded since it usually depends on the laser induced microbubble formation for dispersion or solution samples [44, 45]. Usually, RSA is the dominant mechanism for nonlinear absorption under resonant or near resonant excitation, while the TPA dominates under non-resonant excitation [46, 47]. In our case, the incident laser of 800 nm is far from the resonant wavelength (~1771 nm, corresponding to an optical bandgap of 0.7 eV) of the PdSe$_2$ film. Therefore, the optical limiting behaviour is mainly attributed to TPA. At low laser intensities, the linear absorption is dominant, while as the laser intensity increases, increased photon population makes TPA possible, resulting in the excitation of electrons from the valence to conduction bands.

To extract the TPA coefficient $\beta$ of PdSe$_2$, we fit the measured OA results with Equation (1). The TPA coefficient $\beta$ for the PdSe$_2$ film is shown in Fig. 4(b) at different laser intensities. A large $\beta = 3.26 \times 10^{-8}$ m/W is observed, which is comparable to the reported values of graphene, and higher than that of $WS_2$, highlighting the strong optical limiting effect in $PdSe_2$ film. In addition, the TPA coefficient $β$ is relatively constant with incident laser intensity, reflecting the fact that we are working in an intensity regime where the material properties of the $PdSe_2$ films are not changing much. The slight fluctuation in $β$ with laser intensity may arise from light scattering in the $PdSe_2$ film surface.

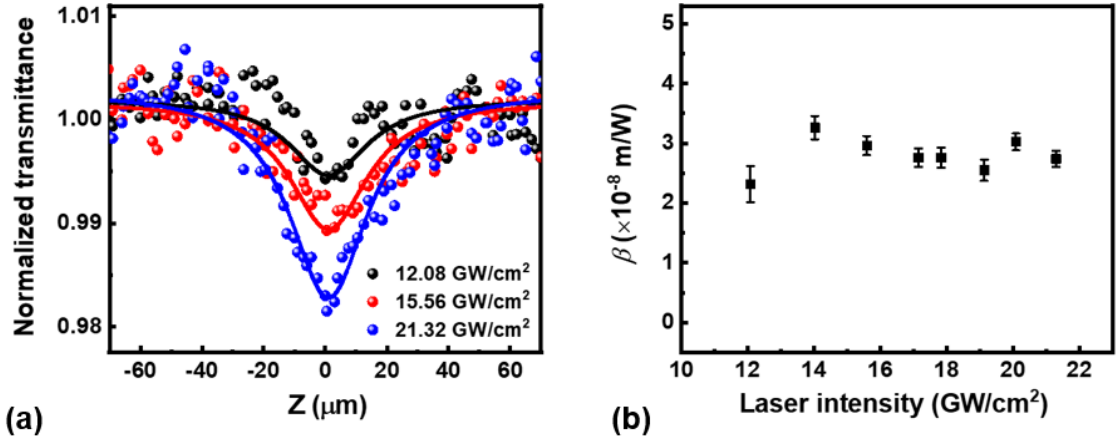

**FIG. 4.** (a) OA Z-scan results of $PdSe_2$ film at different intensities. (b) TPA coefficient $β$ of $PdSe_2$ film versus laser intensity.

To further investigate the nonlinear absorption of the $PdSe_2$ film, we measured the minimum transmittance with the sample at the focal point of the Z scan setup, for different incident laser intensities. Figure 5(a) shows the transmittance of $PdSe_2$ at the focal point as a function of laser intensity, where the transmittance fluctuates around a relatively constant value at low intensities and then decreases significantly as the laser intensity increased. The experimental data fits the theory well [45], verifying our assumption of TPA being the dominant process for nonlinear absorption in the $PdSe_2$ film. The order of the observed nonlinear absorption can also be confirmed by

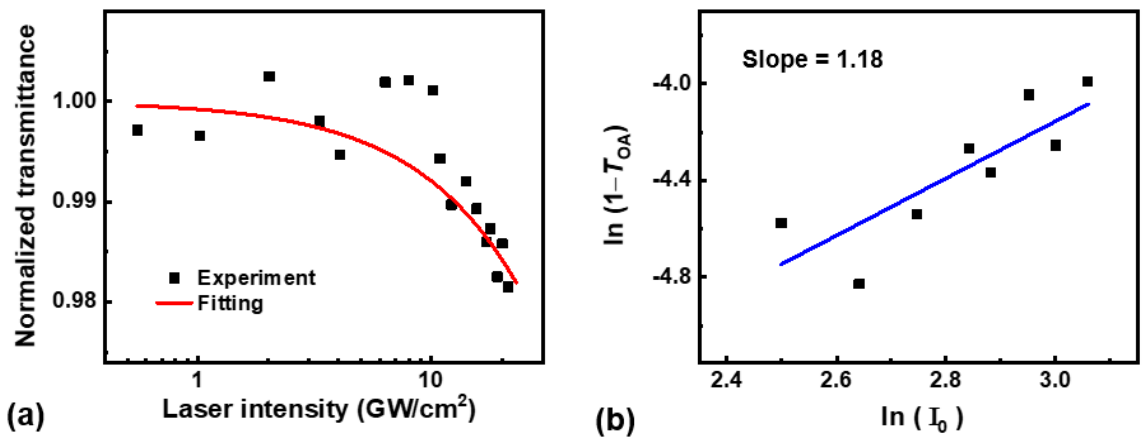

**FIG. 5.** (a) Normalized transmittance of $PdSe_2$ film at the focal point as a function of laser intensity. (b) The plot of $\ln(1−T_{OA})$ versus $\ln(I_0)$ to determine the order of nonlinearity. The measured and fit results are shown by scatter data points and solid lines, respectively.

examining the relation between $\ln(1−T_{OA})$ versus $\ln(I_0)$ [48]:

$$\ln(1 - T_{OA}) = k\ln(I_0) + C, \tag{3}$$

where $k$ is the slope showing the order of the nonlinear absorption and $C$ is a constant. For pure TPA, the slope is equal to 1 [48, 49]. We obtain a slope of 1.18 (Fig. 5(b)), suggesting the observed nonlinear absorption is mainly attributed to TPA in the PdSe$_2$ film.

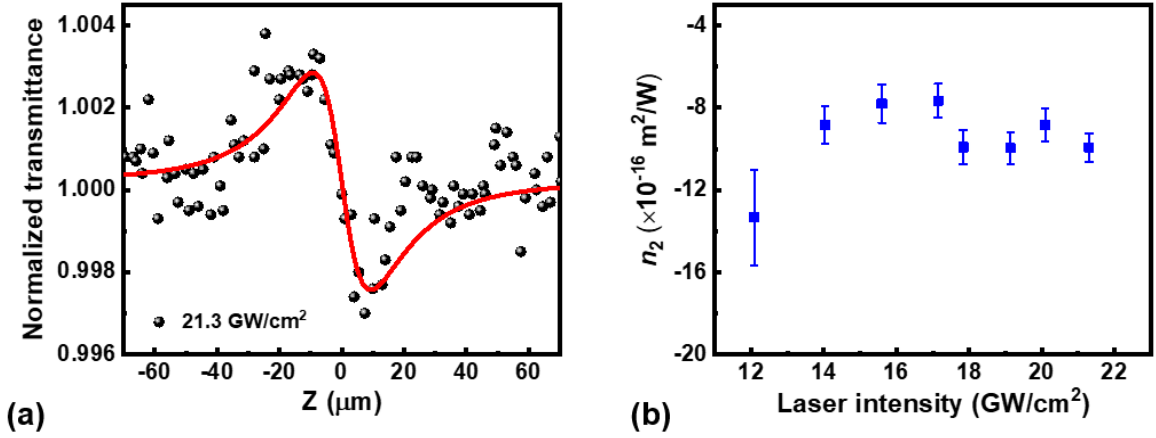

**FIG. 6** (a) CA Z-scan result of PdSe$_2$ film at intensity of 21.32 GW/cm$^2$. (b) Kerr coefficient $n_2$ of PdSe$_2$ film versus laser intensity.

We also performed CA Z-scan measurements to investigate the Kerr nonlinearity ($n_2$) of the PdSe$_2$ films. The values of $n_2$ for the PdSe$_2$ film at different laser intensities were extracted by fitting the measured CA results with Equation 2 [41, 42]. Figure 6(a) shows a representative CA result for PdSe$_2$ at a laser intensity of 21.32 GW/cm$^2$. The transmittance of the sample exhibited a transition from peak to valley when the sample passed through the focal plane. Such a peak-valley CA behaviour corresponds to a negative Kerr coefficient $n_2$ and indicates an optical self-defocusing effect in the PdSe$_2$ film. As discussed above, TPA results in the transfer of electrons from valence band to conduction band, increasing the free carrier density in the film. Therefore, the observed negative Kerr nonlinearity potentially originates from the TPA-induced free carrier nonlinear refraction and interband blocking [50-52].

Fig. 6(b) shows the measured Kerr coefficient $n_2$ of PdSe$_2$ versus laser intensity, showing a large $n_2$ of $\sim -1.33 \times 10^{-15}$ m$^2$/W. Table 1 compares the $\beta$ and $n_2$ of PdSe$_2$ with other 2D layered materials. As can be seen, the value of $n_2$ for PdSe$_2$ is lower than those of graphene and GO, but still more than two orders of magnitude higher than bulk silicon [53, 54]. Such a high $n_2$ suggests that PdSe$_2$ is an extremely promising material for self-defocusing based nonlinear photonic applications. For example, a negative Kerr nonlinearity can be used to self-compress ultrashort pulses in the presence of positive group-velocity dispersion [55, 56]. Another application of a negative Kerr nonlinearity is mode locking of lasers using the Kerr mode-locking technique [53, 57] as well as the possibility of achieving net parametric modulational instability gain under normal dispersion conditions [53, 58].

Moreover, as shown in Fig. 6(b), the absolute value of $n_2$ initially decreases with laser intensity and then saturates at higher intensities. In theory, the optical nonlinear refraction originates mainly from the free-carrier and bound-electron nonlinearities [51, 59, 60]. We assume that both mechanisms exist in PdSe$_2$ films. The refractive index change in the PdSe$_2$ film can be expressed by $\Delta n = n_2^* I_0 + \sigma_r N$, where $n_2^*$ is the nonlinear refraction related to bonding electrons, $\sigma_r$ is the free carrier refractive coefficient and $N$ is the charge carrier density.[51] Therefore, the effective $n_2 = \Delta n/I_0 = n_2^* + \sigma_r N/I_0$, is an intensity dependent parameter, which can explain the $n_2$ variation as a function of laser intensity observed in our measurements.

Table 1. Comparison of $\beta$ and $n_2$ for various 2D layered materials

| Material | Laser parameter | Thickness | $\beta$ (m/W) | $n_2$ (m$^2$/W) | $n_2$ (×$n_2$ of Si[1]) | Ref. |
|---|---|---|---|---|---|---|
| Graphene | 1150 nm, 100 fs | 5-7 layers | $3.8 \times 10^{-8}$ | $-5.5 \times 10^{-14}$ | $-1.22 \times 10^4$ | [61] |
| GO | 800 nm, 100 fs | ~2 μm | $4 \times 10^{-7}$ | $1.25 \times 10^{-13}$ | $2.75 \times 10^4$ | [62] |
| MoS$_2$ | 1064 nm, 25 ps | ~25 μm | $(-3.8 \pm 0.59) \times 10^{-11}$ | $(1.88 \pm 0.48) \times 10^{-16}$ | 41.32 | [18] |
| WS$_2$ | 1040 nm, 340 fs | ~57.9 nm | $(1.81 \pm 0.08) \times 10^{-8}$ | $(-3.36 \pm 0.27) \times 10^{-16}$ | -73.85 | [19] |
| BP | 1030 nm, 140 fs | ~15 nm | $5.845 \times 10^{-6}$ | $-1.635 \times 10^{-12}$ | $-3.59 \times 10^5$ | [14] |
| PtSe$_2$ | 800 nm, 150 fs | ~4.6 nm | $-8.80 \times 10^{-8}$ | – | – | [63] |
| PtSe$_2$ | 1030 nm, 340 fs | 17 layers | – | $(-3.76 \pm 0.46) \times 10^{-15}$ | $-8.26 \times 10^2$ | [48] |
| PdSe$_2$ | 800 nm, 140 fs | ~8 nm | $(3.26 \pm 0.19) \times 10^{-8}$ | $(-1.33 \pm 0.23) \times 10^{-15}$ | $-2.92 \times 10^2$ | This work |

[1]$n_2$ for silicon = $4.55 \times 10^{-18}$ m$^2$/W (ref. [54])

In contrasting the Kerr 3$^{rd}$ order nonlinearity and TPA with other 3$^{rd}$ order effects such as optical third harmonic generation (THG) [64 - 69] it is interesting to note that unlike THG, both the sign and phase of the Kerr nonlinearity are critically important. The concept of the nonlinear FOM was originally proposed purely in the context of a limitation for $n_2$ devices having a positive $n_2$ and TPA. Here, the fact that $n_2$ for PdSe$_2$ is negative means that the conventional importance of a large FOM is probably not applicable. Further, TPA has been known to potentially have a positive impact on some signal processing functions [70 - 72] and so the large TPA of PdSe$_2$ is a potential advantage of this material. The combination of a large and negative $n_2$ with a very large TPA makes PdSe$_2$ a highly interesting material for novel nonlinear processes that can exploit self-defocusing effects.

## 4. Conclusion

We report a large third-order nonlinear optical response of PdSe$_2$ films measured with the Z-scan technique. Our results show that PdSe$_2$ has a strong TPA response with a large $\beta$ of ~ 3.26 ×10$^{-8}$ m/W. We also investigate the Kerr nonlinearity ($n_2$) of PdSe$_2$ finding that $n_2$ is negative, and with an absolute magnitude that is more than two orders of magnitude larger than bulk silicon. Furthermore, the variation in $n_2$ of PdSe$_2$ with laser intensity is characterized. We find that $n_2$ initially increases (decreasing in absolute magnitude) with incident laser intensity and then saturates at higher intensities. These results verify PdSe$_2$ as a promising 2D material with prominent nonlinear optical properties.

## Acknowledgements

This work was supported by the Australian Research Council Discovery Projects Program (No. DP150102972 and DP190103186), and the Industrial Transformation Training Centres scheme (Grant No. IC180100005). We acknowledge Swinburne Nano Lab and Micro Nano Research Facility (MNRF) of RMIT University for the support in material characterization as well as Shenzhen Sixcarbon Technology for the PdSe$_2$ film fabrication. We thank Dr. Yunyi Yang and Dr. Tania Moein for technical support, Dr. Deming Zhu for assisting in XPS characterization and Dr. Chenglong Xu for assisting in optical characterization.